\documentclass[10pt,letterpaper]{article}
\usepackage{subcaption}
\usepackage{opex3}
\usepackage{epstopdf}
\usepackage{graphicx}
\usepackage{cite}

\begin{document}
\title{Multilayer nanoparticle arrays for broad spectrum absorption enhancement in thin film solar cells}

\author{Aravind Krishnan,$^{1,*}$ Snehal Das,$^{1}$ Siva Rama Krishna,$^{1}$ and Mohammed Zafar Ali Khan$^{1}$}
 
\address{$^{1}$ Department of Electrical Engineering, Indian Institute of Technology Hyderabad, ODF Estate, Medak District, Andhra Pradesh-502205, India}

\email{$^{*}$ee10b004@iith.ac.in} 

\begin{abstract} In this paper, we present a theoretical study on the absorption efficiency enhancement of a thin film amorphous Silicon (a-Si) photovoltaic cell over a broad spectrum of wavelengths using multiple nanoparticle arrays. The light absorption efficiency is enhanced in the lower wavelengths by a nanoparticle array on the surface and in the higher wavelengths by another nanoparticle array embedded in the active region. The efficiency at intermediate wavelengths is enhanced by the simultaneous resonance from both nanoparticle layers. We optimize this design by tuning the radius of particles in both arrays, the period of the array and the distance between the two arrays. The optimization results in a total quantum efficiency of $62.35\%$ for a 0.3 $\mu m$ thick a-Si substrate.\end{abstract}

\ocis{250.5403; 240.6680; 250.0250; 310.6845; 350.6050; 290.4020.}



\section{Introduction}
Thin film solar cells have become a potential alternative to traditional crystalline silicon solar cells due to lower manufacturing costs, better flexibility, durability and shorter energy payback period. A better understanding of the underlying physics and the availability of mature fabrication technology makes amorphous silicon the preferred choice for thin film photovoltaic cells. But the widespread adoption of thin film solar cells is limited by its relatively poor absorption efficiency owing to the indirect band gap of silicon~\cite{Atwater}. In general, the efficiency of solar cells can be improved either by increasing the light absorption efficiency or by increasing the minority carrier life time. In order to increase the light absorption, it becomes necessary to incorporate light trapping mechanisms in solar cells. Recent advancements in the field of plasmonics have opened up the possibility of using metallic nanoparticles to increase the absorption of thin film solar cells~\cite{Atwater,Catchpole,Akimov}. Nanoparticles exhibit the phenomenon of surface plasmon resonance when illuminated with light of suitable frequency~\cite{Pillai,Barnes}. Near the plasmon resonance frequency, metallic nanoparticles strongly scatter light incident on it. This scattering effect is the result of the collective oscillations of electrons to re-radiate electromagnetic radiation~\cite{Bohren83}. When placed on the surface of silicon substrate, metallic nanoparticles scatter light preferentially into it due to the high refractive index of silicon~\cite{Spinelli}. The particles also scatter the incident light at an angle, thereby increasing the path length of photons within the silicon substrate. As a result, the optical thickness of the active region increases, thereby allowing the semiconductor substrate to absorb higher levels of electromagnetic radiation~\cite{Atwater}. 

Researchers have explored the possibility of placing nanoparticles on the surface~\cite{Akimov,Pillai,Chen}, rear~\cite{Pillai11} and within the active region~\cite{Choi} of thin film solar cells. Multiple arrays of nano-structures placed either on the front, rear or both are used to further enhance the absorption efficiency~\cite{Ho,Shi,Sefunc}. The phenomenon of energy transfer between discrete nano-structures located at the surface and active region of a solar cell is used to control the flow of energy and improve absorption efficiency~\cite{Lin}. In these cases, enhancing the quantum efficiency has been considered over limited bands of wavelengths. Also, a loss in performance is possible if the metallic nanoparticles are introduced in the bulk due to increased recombination~\cite{Santbergen,Xue}. However, it is possible to prevent the metallic nanoparticles from acting as recombination centers either by using room temperature processes such as microcontact printing~\cite{Santhanam} or by passivating it with a dielectric coating~\cite{Choi}. Motivated by these facts, in this paper, we enhance the light absorption efficiency of a thin film a-Si solar cell by placing periodic arrays of silver nanoparticles at the surface and within the active region of the cell. The surface and bulk layer of nanoparticles enhances the absorption of lower and higher wavelengths respectively. In addition the absorption of light within the substrate is further enhanced by the simultaneous resonance from the surface and bulk layers at intermediate wavelengths. Due to the multiple and high-angle scattering from both the layers of nanoparticles, the effective optical path length of light increases inside the cell. This results in several fold increase in the optical thickness of the active region without increasing the physical thickness~\cite{Atwater}.

\section{Background theory}
Understanding the basic mechanism of localized surface plasmon resonance and various parameters affecting it is necessary to engineer the nanoparticles for light trapping applications in solar cells. The resonant behavior of localized surface plasmons are restricted to a limited range of frequencies determined by the size and shape of the particles, dielectric functions of the involved media, and the electromagnetic interaction between them~\cite{Kreibig}. When the frequency of light incident on the nanoparticles match the natural frequency of the oscillating surface electrons, the phenomenon of surface plasmon resonance is established~\cite{Zeng}. The collective response of the electrons in spherical metallic nanoparticles with dimension much lesser than the wavelength of light is described by the dipolar polarizability $\alpha$;~\cite{Maier}

\begin{equation}
\alpha = 3V \left(\frac{\epsilon - \epsilon_m}{\epsilon + 2\epsilon_m}\right)
\end{equation}

where V is the particle volume, $\epsilon_m$ is the dielectric function of the embedding medium and $\epsilon$ is the dielectric function of the metallic nanoparticle. The size dependent dielectric function the particle $\epsilon(\omega,D)$ is given by,~\cite{Kreibig}

\begin{equation}
\epsilon(\omega,D) = \epsilon_{IB}+\left(1- \frac{\omega_p^2}{\omega^2 +i \omega \gamma(D)}\right)
\end{equation}

where $D$ is the diameter of the particle and $i=\sqrt{-1}$. The size independent first term $\epsilon_{IB}$ is due to interband transitions. The second term is the Drude-Sommerfeld free electron term which contains the bulk plasmon frequency $\omega_p$, given by $\omega_p^2=Ne^2/m\epsilon_0$ where N is the density of free electrons, e is the electronic charge, m is the effective mass of an electron and $\epsilon_0$ is the free-space dielectric constant, and the size-dependent damping constant $\gamma$;\cite{Kreibig}

\begin{equation}
 \gamma(D)=\gamma_0+2\left(\frac{Av_F}{D}\right)
\end{equation}

where $\gamma_0$ is the bulk damping rate and $v_F$ is the Fermi velocity of the electrons. A is a dimensionless size parameter which accounts for the additional surface damping terms, such as the inelastic collisions between electrons and chemical interface damping. Inelastic collisions shorten the mean free path of the electrons when nanoparticle size decreases. Chemical interface damping occurs in embedded nanoparticles, where the fast energy transfer between the particle and its surroundings results in the loss of phase coherence of the collective electron oscillation.

The particle polarizability becomes large when $\epsilon = -2\epsilon_m$, and this effect is called surface plasmon resonance. At resonance the particle scattering cross-section area, $C_{scat}$, a parameter that determines the extent to which scattering of light occurs, is several times the geometric cross section of the particle~\cite{Bohren83}.

\begin{equation}
C_{scat} = \frac{1}{6\pi}\left(\frac{2\pi}{\lambda}\right)^4 |\alpha|^2
\end{equation}

where $\lambda$ is the wavelength of the incident light. For particles with diameter in the range of a few nanometers, scattering is also accompanied by absorption. As particle diameter increases to around 100 nanometers, the scattering effect strongly outweighs the absorption by the particle~\cite{Catchpole}. Also, the resonance is accompanied by another process, namely dynamic depolarization~\cite{Meier}. As size increases, the oscillation of the conduction electrons at resonance goes out of phase, resulting in a red shift of the particle resonance. The increase in size also results in higher radiation damping, which results in the broadening of the range of frequencies at which resonance occurs~\cite{Kottman,Dahmen}. The surface plasmon resonance can be tuned to a large extent either by coating or sandwiching the nanostructure with a dielectric medium. The medium can markedly enhance the red shift of the central wavelength and consequently obtain a wide tunable range at higher wavelengths~\cite{Xu}. Both red shifting and broadening of resonance frequencies are advantageous for the solar cell~\cite{Catchpole}. 
  
\section{Proposed Solar Cell Design}
In our thin film solar cell design, the active region is made of amorphous silicon having a thickness of $0.3\;\mu m$.  Spherically shaped silver nanoparticles are placed periodically in rectangular arrays on the surface as well as inside the active region of the substrate. Particle layers are placed in such a way that nanoparticles in both layers are vertically aligned. The perspective view of the design described above is shown in Fig. 1(a). The schematic diagram in Fig. 1(b) shows the various parameters of the proposed design. The radius of the nanoparticles on the surface and in the active region are denoted by $R_s$ and $R_b$ respectively. The period of the nanoparticle array is denoted by $P$ and the distance from the surface of the solar cell to the top of the bulk array is denoted by $T$. 

\begin{figure}[ht]
\centering
\begin{subfigure}{.48\textwidth}
  \centering
  \includegraphics[width=1\linewidth]{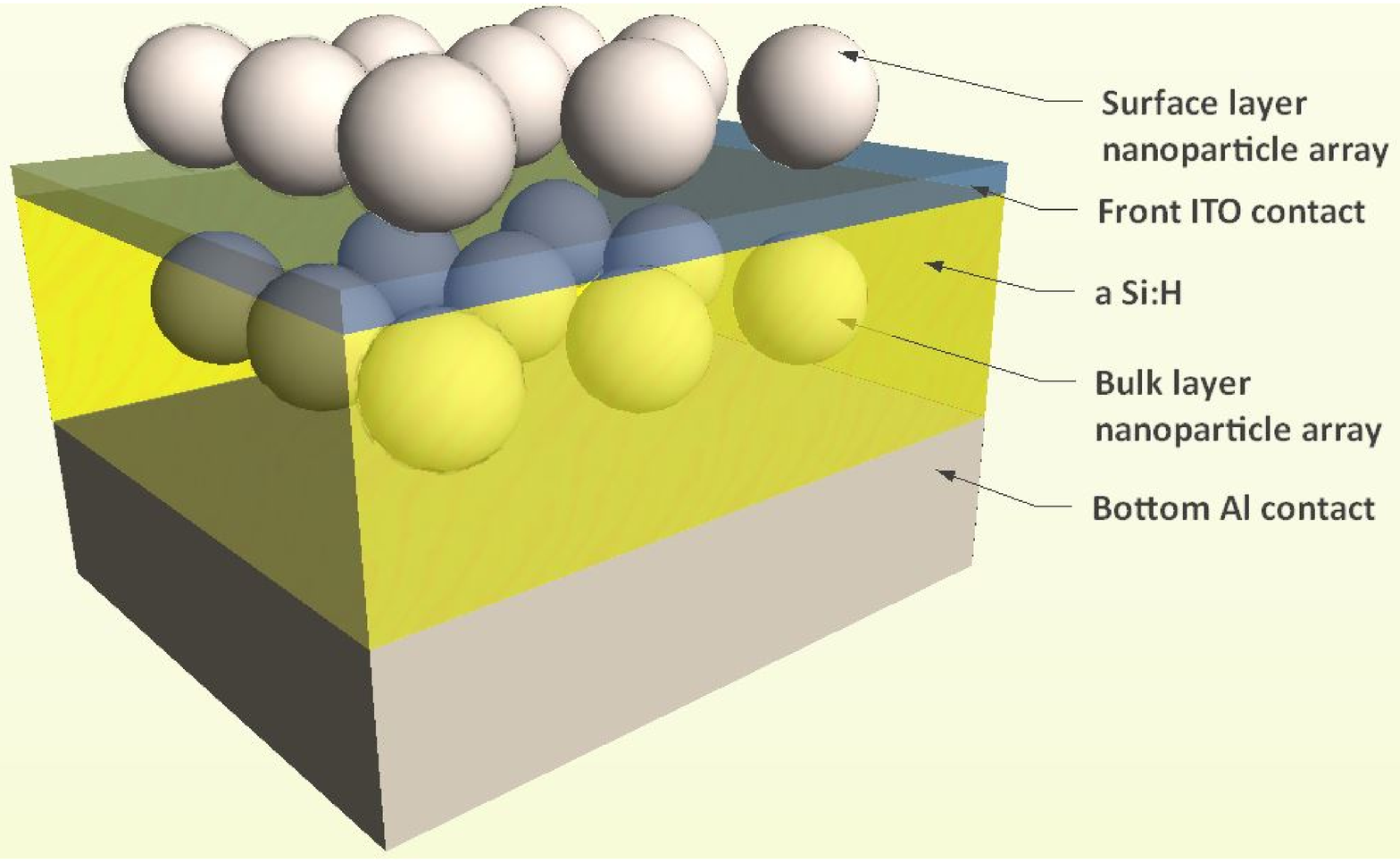}
  \caption{}
\end{subfigure}
\begin{subfigure}{.48\textwidth}
  \centering
  \includegraphics[width=0.8\linewidth]{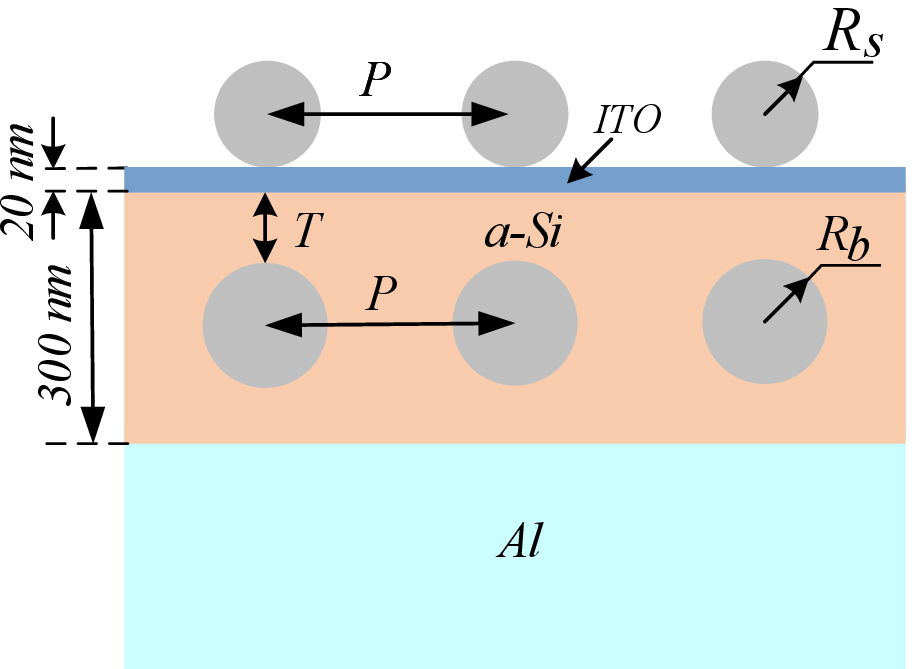}
  \caption{}
\end{subfigure}
\caption{(a) Perspective view of the solar cell with silver nanoparticle arrays on the surface and in the bulk.(b) Various parameters of the considered design.
}
\end{figure}

The device performance is compared with a few existing reference designs to understand the improvement in absorption efficiency. A $300\;nm$ thick active region of amorphous silicon with an Aluminium back contact and a $20~nm$ thick ITO front contact is considered for all designs~\cite{Zhao,Deng,Akimov_Koh,Wu}. Design 1, considered as the primary reference, contains only the a-Si layer and does not incorporate any nanoparticle arrays for light trapping. Design 2 contains silver nanoparticle arrays on top of the ITO layer, similar to the design in~\cite{Akimov_Koh}. A silver nanoparticle array is introduced into the bulk of the amorphous silicon in Design 3. Design 4 corresponds to the proposed structure where nanoparticle arrays are introduced at both the surface and bulk of the device. Initially the radii of the nanoparticles on the surface ($R_s$) and in the active region ($R_b$) are chosen as $50\;nm$ and $65\;nm$ respectively. The purpose is to exploit the plasmonic enhancement provided by different layers at different wavelength regimes so as to enhance the absorption over a wide band of wavelengths. The nanoparticles on the surface resonates at lower wavelengths and provides plasmonic enhancement in these regions. On the other hand the particle layer in the bulk with nanoparticles of higher size resonates at higher wavelengths and provide plasmonic enhancement in these wavelengths. In order to allow horizontal and vertical coupling of light between the nanoparticles, the period of the nanoparticle arrays ($P$) is taken as $200\;nm$ and the distance from the surface of the solar cell to the top of the bulk array ($T$) is taken as $85\;nm$. These parameters are optimized in section-5 to determine the best possible configuration that makes the maximum absorption enhancement. The initial specifications of the nanoparticle arrays for designs 2 to 4 are summarized in Table 1.

\begin{table}[h]
{\bf \caption{Specifications for the proposed design and reference designs (all dimensions are in $nm$)}}
\begin{center}
\begin{tabular}{lllll}\hline
Design &  $R_s$ &  $R_b$ & $P$ & $T$ \\ \hline
Design 1 & - & - & - & - \\
Design 2 & 50 & - & 200 & - \\
Design 3 & - & 65 & 200 & 85 \\
Design 4 & 50 & 65 & 200 & 85 \\ \hline
\end{tabular}
\end{center}
\end{table}

The three dimensional finite difference time domain analysis (FDTD) tool provided by Lumerical is used for the design simulation. The amorphous silicon material is optically modeled using the refractive index data provided in the SOPRA N$\&$K Database~\cite{SOPRA}. A normally incident plane wave source of unit amplitude with a wavelength range from $400\;nm$ to $1100\;nm$ is placed above the surface nanoparticle layer. Periodic boundary conditions are used for the side boundaries to model the periodic nature of the particles. The Perfectly Matched Layer (PML) boundary conditions are used for upper and lower boundary to approximate the effect of infinite space and infinite bottom contact respectively. Mesh size of $2~nm$ is used for the regions where the particles are present and $3~nm$ for the a-Si absorber region. The absorbed power within the active region of the solar cell is calculated with the help of 3-D power monitors placed on the a-Si substrate. The monitors record field data over the entire volume of the substrate and find the power absorbed as a function of spatial co-ordinates. For the design with particles embedded in the bulk, the power absorbed by the particles has to be deducted from the net absorbed power, since it does not contribute to the electron-hole pair generation. The power absorbed by the silicon absorber layer is calculated by applying a spatial filter which integrates the absorption within the filter region (here the a-Si substrate). The quantum efficiency and total quantum efficiency are calculated for all the designs considered~\cite{Shi}.

The quantum efficiency of a solar cell for a particular wavelength, $ QE(\lambda) $, is defined as the ratio of the absorbed power within the solar cell, $ P_{abs}(\lambda) $, to the incident power of light, $P_{in}(\lambda)$;

\begin{equation}
QE(\lambda) = \frac{P_{abs}(\lambda)}{P_{in}(\lambda)}.
\end{equation}

The quantum efficiency of all designs in Table 1 are compared in Fig. 2 to understand the improvement in absorption efficiency of the proposed design at different wavelengths.

\begin{figure}[htb]
\centerline{\includegraphics[width=8.3 cm]{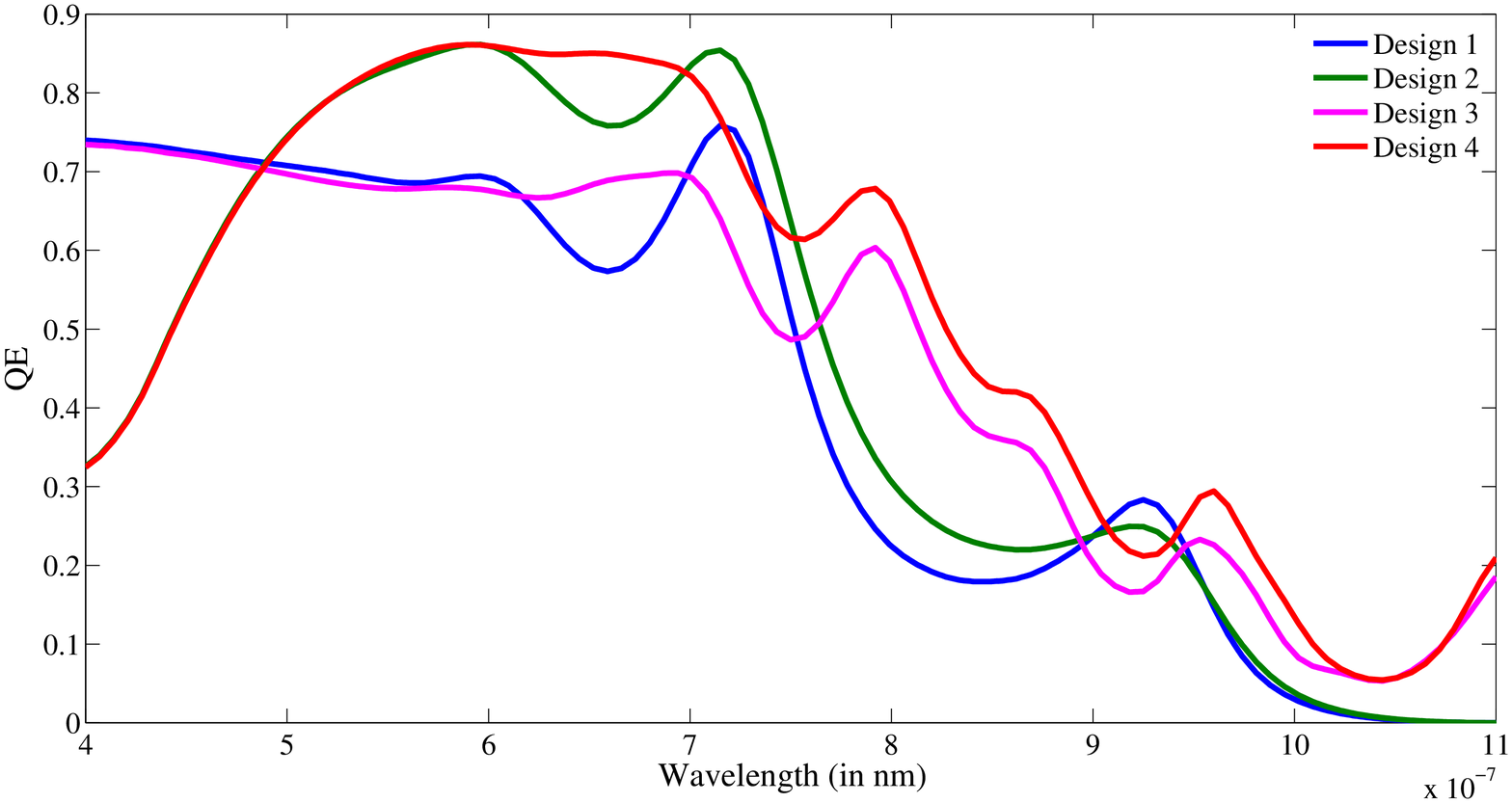}}
\caption{Quantum efficiency of various designs under consideration as a function of wavelength of light.}
\end{figure}

It can be observed from Fig. 2 that the design that employs nanoparticles on the surface and active region of the solar cell shows significant improvement in absorption efficiency over the reference designs. The introduction of nanoparticles on the surface improves the absorption efficiency of a-Si for a range of wavelengths between $500\;nm$ and $900\;nm$, but causes little improvement at higher wavelengths. On the other hand, the introduction of nanoparticles inside the active region results in significant improvement in the absorption efficiency at most wavelengths above $620\;nm$, even though it has minimal effect at lower wavelengths. The proposed design exploits the absorption enhancement caused by both layers and produces improved absorption at all wavelengths. Around $620-900\;nm$, the simultaneous resonance from particles in both the layers enhances the field in the active region. The inter-layer particle interaction produces an improvement in photon absorption when compared to the enhancements caused by individual layers. The decrease in absorption efficiency below $500~nm$ for designs 2 and 4 is due to the screening effect caused by the nanoparticle layer on the surface of the device. The plasmonic enhancement of the surface layer particles is negligible at these wavelengths and it prevents most of the incoming light from penetrating into the substrate. The introduction of nanoparticles on the bulk reduces the enhancement due to bottom metallic contact which results in a superior performance for Design 1 and 2 over the proposed device around wavelengths $720~nm$ and $940~nm$. The field profile inside the amorphous silicon substrate at various wavelengths is given in Fig. 3. 

\begin{figure}[h]
\centerline{\includegraphics[width=13 cm]{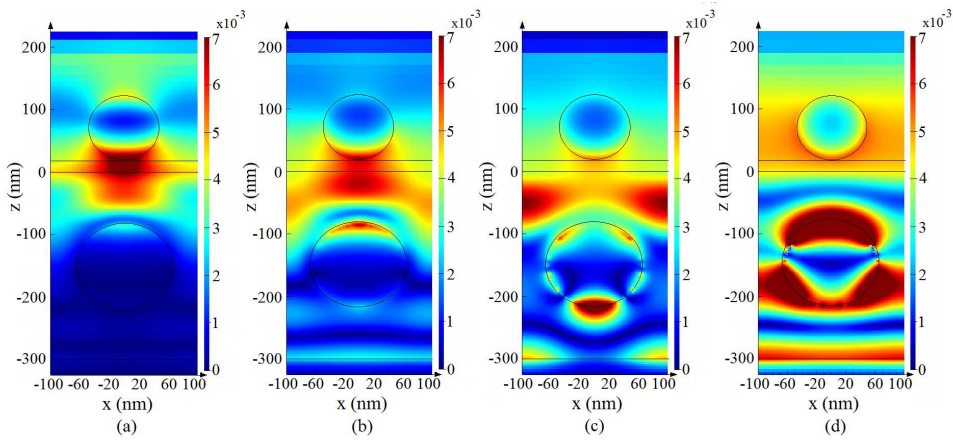}}
\caption{Field profile at different wavelengths inside the amorphous silicon substrate of Design 4. Field Distribution corresponds to wavelengths $547\;nm$, $666~nm$, $799\;nm$ and $1030\;nm$ respectively are shown from left to right.}
\end{figure}

It is evident from Fig. 3(a) that the high energy radiation is absorbed near the surface and fails to penetrate enough into the substrate so as to resonate the nanoparticles buried inside the amorphous silicon. At higher wavelengths, silicon has poor absorption, as a result of which the incident radiation penetrates deep into the substrate, as seen in Fig. 3(d). This is in accordance with Beer-Lambert's Law, which states that the extent of absorption in a solid depends inversely on the wavelength of the incident radiation~\cite{Swinehart}. In order to enhance the light absorption in lower wavelengths, particles with a surface plasmon resonance frequency falling in this wavelength region are placed on the surface of the active region. Larger particles are employed in the active region of the substrate for providing the plasmonic enhancement at higher wavelengths, since an increase in particle size and embedding dielectric medium causes a red shift in the resonance frequency. The radii of particles in both the layers are chosen in such a way that their resonance frequency does not fall in very low or very high frequencies. This is done so as to exploit the enhancement due simultaneous resonance from both the nanoparticle layers at intermediate wavelengths. The field distribution at wavelengths $666~nm$ and $799~nm$ shown in Fig. 3(b) and Fig. 3(c) respectively depicts the field enhancement due to this simultaneous resonance of nanoparticle layers. 


For the broad spectrum of incident electromagnetic radiation, the total quantum efficiency, $TQE$, which takes the solar spectral irradiance into account, determines the overall absorption efficiency of the solar cell. $TQE$ is the fraction of incident photons that are absorbed by the solar cell.

\begin{equation}
TQE = \frac{\int_{\lambda_1}^{\lambda_2} \frac{\lambda}{hc} QE(\lambda) I_{AM1.5}(\lambda)\,\mathrm{d}\lambda}{\int_{\lambda_1}^{\lambda_2} \frac{\lambda}{hc} I_{AM1.5}(\lambda)\,\mathrm{d}\lambda}
\end{equation}

where \textit{h} is Planck's constant, \textit{c} is the speed of light in the free space and $I_{AM1.5}$ is the AM 1.5G solar spectrum. $TQE$ and absorption enhancement for the broad spectrum of light as compared to the plain reference silicon solar cell are listed in Table 2.
\begin{table}[h]
{\bf \caption{Comparison of $TQE$ of the proposed design with the other considered designs and their enhancement as compared to a-Si without nanoparticles}} 
\begin{center}
\begin{tabular}{lll}\hline
Design & $TQE$ (\%) & enhancement \\ \hline
Design 1  & 43.71 & 1.0000 \\
Design 2  & 49.09 & 1.1231 \\
Design 3  & 49.21 & 1.1258 \\
Design 4  & 56.01 & 1.2814 \\ \hline
\end{tabular}
\end{center}
\end{table}

The absorption enhancement over the entire wavelength region translates to a higher $TQE$ for Design 4 when compared to the other considered designs. A $TQE$ of $56.01\%$ is obtained for the proposed design. 

\section{Variation of absorption efficiency with design parameters}
There are various parameters of the proposed design that influence the absorption profile of thin film substrate. Change in nanoparticle radius shifts the surface plasmon resonance frequency whereas the mutual interaction between the nanoparticle layers is affected by a change in particle layer separation as well as the alignment of particles in both the arrays. The effect of these parameters towards the absorption efficiency of the solar cell is studied in detail. 

\subsection{Influence of particle layer separation in absorption efficiency}
The variation in absorption efficiency $(QE(\lambda))$ with change in inter particle layer separation $(T)$ is investigated first. $T$ is varied from $17\;nm$ to $153\;nm$ while keeping the  period of both nanoparticle arrays at $200\;nm$ and radii of nanoparticles in surface and bulk layer at $50\;nm$ and $65\;nm$ respectively (same as in design 4). $QE(\lambda)$ and $TQE$ for various inter layer separations is given in Fig. 4(a) and Fig. 4(b) respectively.

\begin{figure}[ht]
\centering
\begin{subfigure}{.5\textwidth}
  \centering
  \includegraphics[width=1\linewidth]{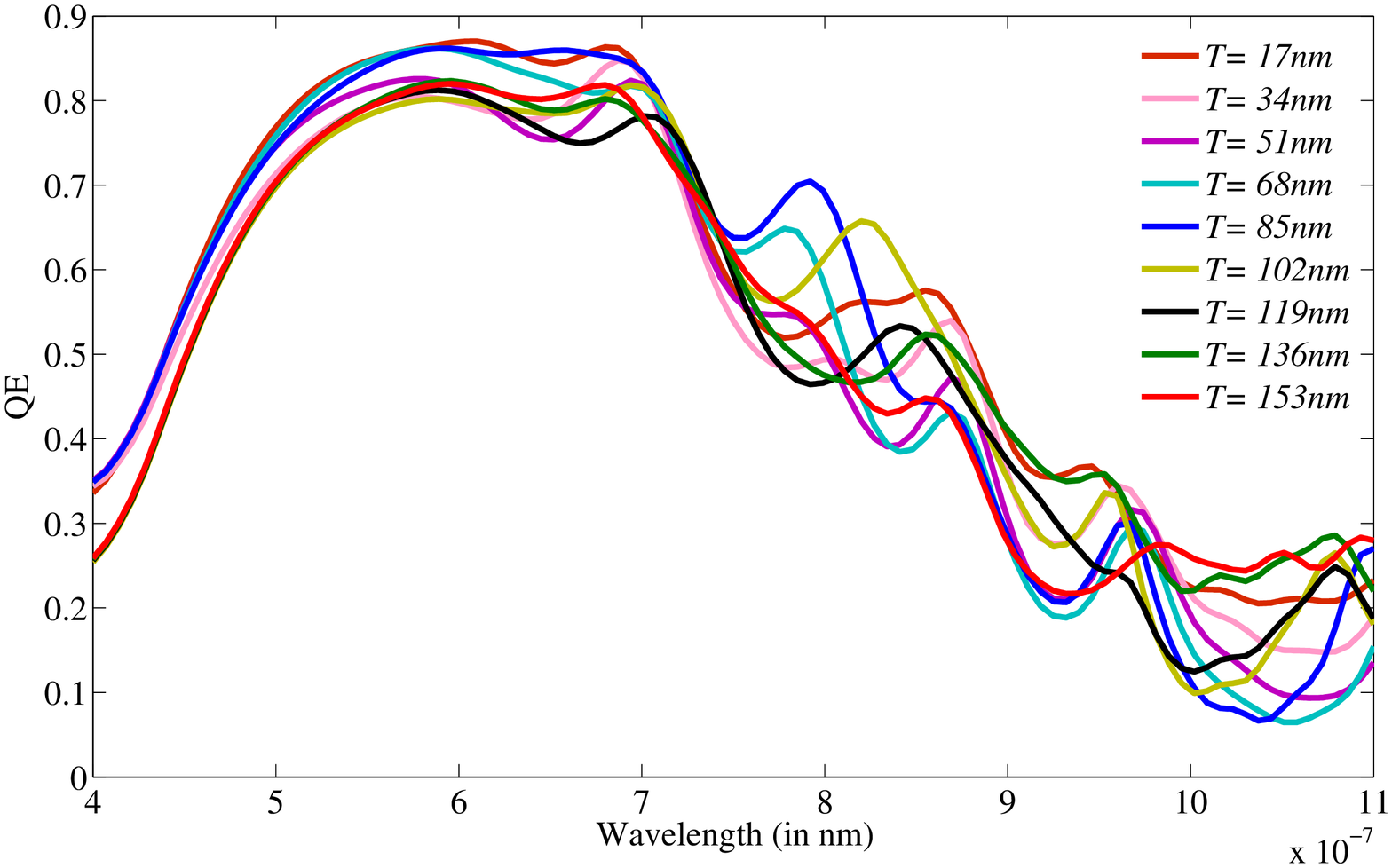}
  \caption{}
\end{subfigure}
\begin{subfigure}{.49\textwidth}
  \centering
  \includegraphics[width=1\linewidth]{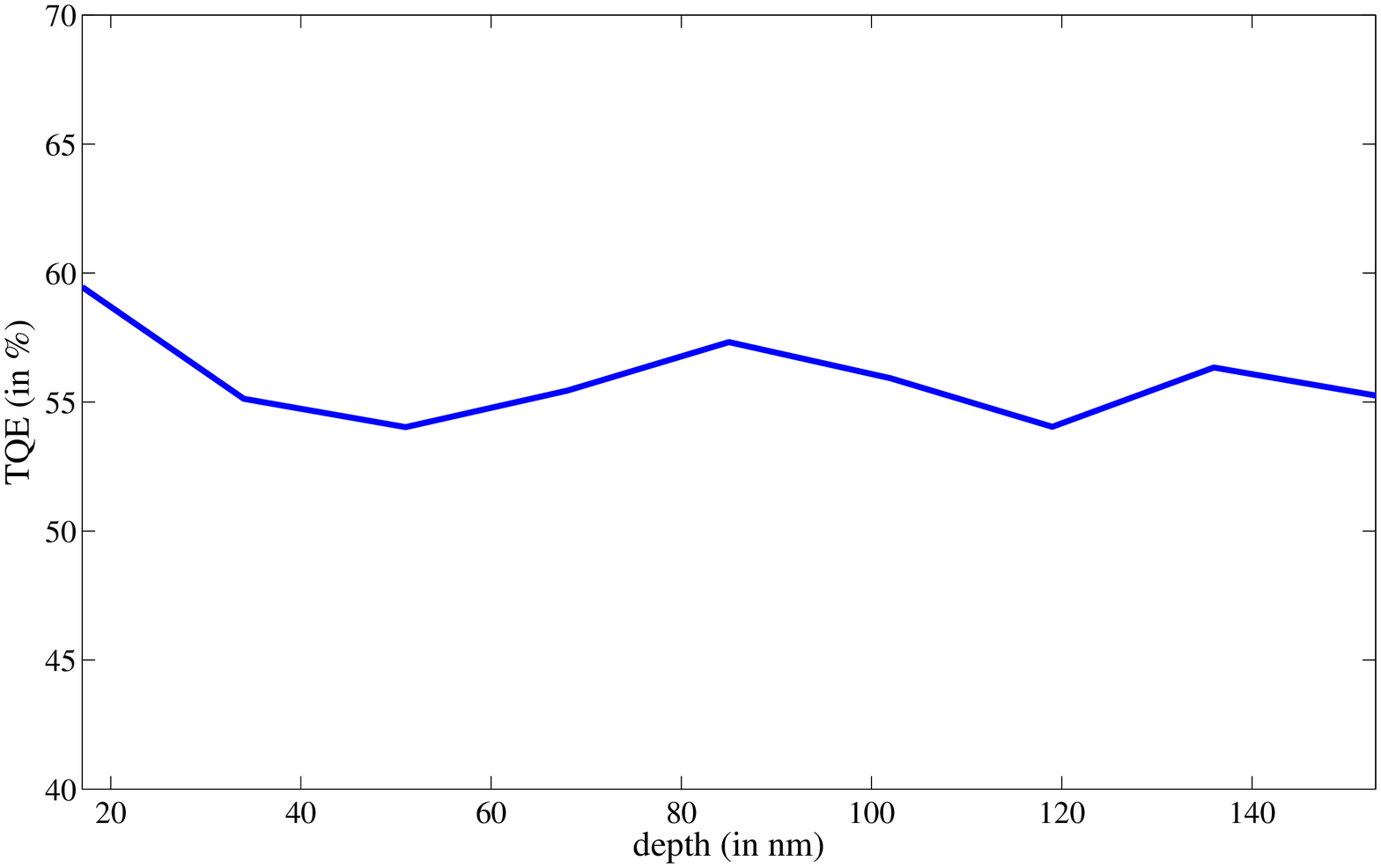}
  \caption{}
\end{subfigure}
\caption{Variation of $QE$ and TQE with change in inter particle layer separation }
\end{figure}

For wavelengths below $700~nm$ a slight improvement in absorption efficiency is observed when the bulk nanoparticle layer is located close to the surface of a-Si substrate. Maximum absorption in the $700-800~nm$ wavelength range occurs when the nanoparticles in the bulk is located in the middle of the active region. For wavelengths above $850~nm$ placing the particles near to the bottom of the substrate produces better enhancement in absorption. The overall effect while considering the entire wavelength region is given by the $TQE$ measurement. The $TQE$ for different particle layer separations is shown in Fig. 4(b). The net absorption is maximum when the nanoparticle layer embedded in the bulk is kept closer to the surface layer. There is a peak in $TQE$ when the bulk nanoparticle layer is located near to the middle of the substrate and towards the bottom of the substrate. This is due to the absorption enhancement at these positions for wavelength range $700-800~nm$ and $850-1100~nm$ respectively.

\subsection{Influence of particle radii in absorption efficiency}
The size of silver nanoparticles in both surface and bulk layer influences the absorption inside sandwiched a-Si substrate. The variation in absorption efficiency of the design with change in nanoparticle radii of surface layer $(R_s)$ and bulk layer $(R_b)$ is studied separately. $R_s$ is varied from $0$ to $100\;nm$ while keeping the other design parameters same as that of Design 4. Similarly $R_b$ is also varied from $0$ to $100\;nm$ with other parameters kept at the same value as in Design 4. The variation of $QE(\lambda)$ with $R_s$ and $R_b$ are shown in Fig. 5(a) and Fig. 5(b) respectively.

\begin{figure}[ht]
\centering
\begin{subfigure}{.49\textwidth}
  \centering
  \includegraphics[width=1\linewidth]{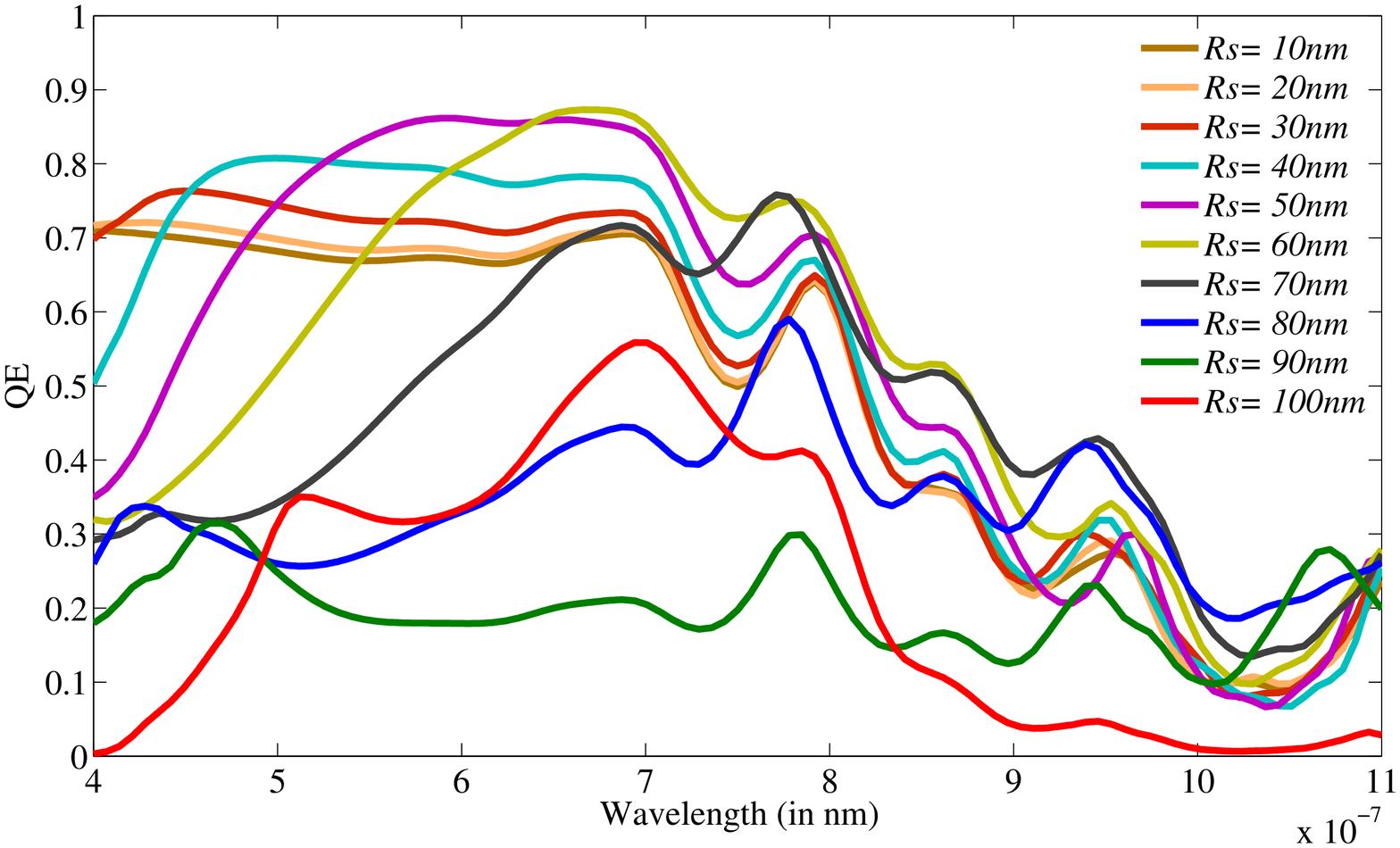}
  \caption{}
\end{subfigure}
\begin{subfigure}{.49\textwidth}
  \centering
  \includegraphics[width=1\linewidth]{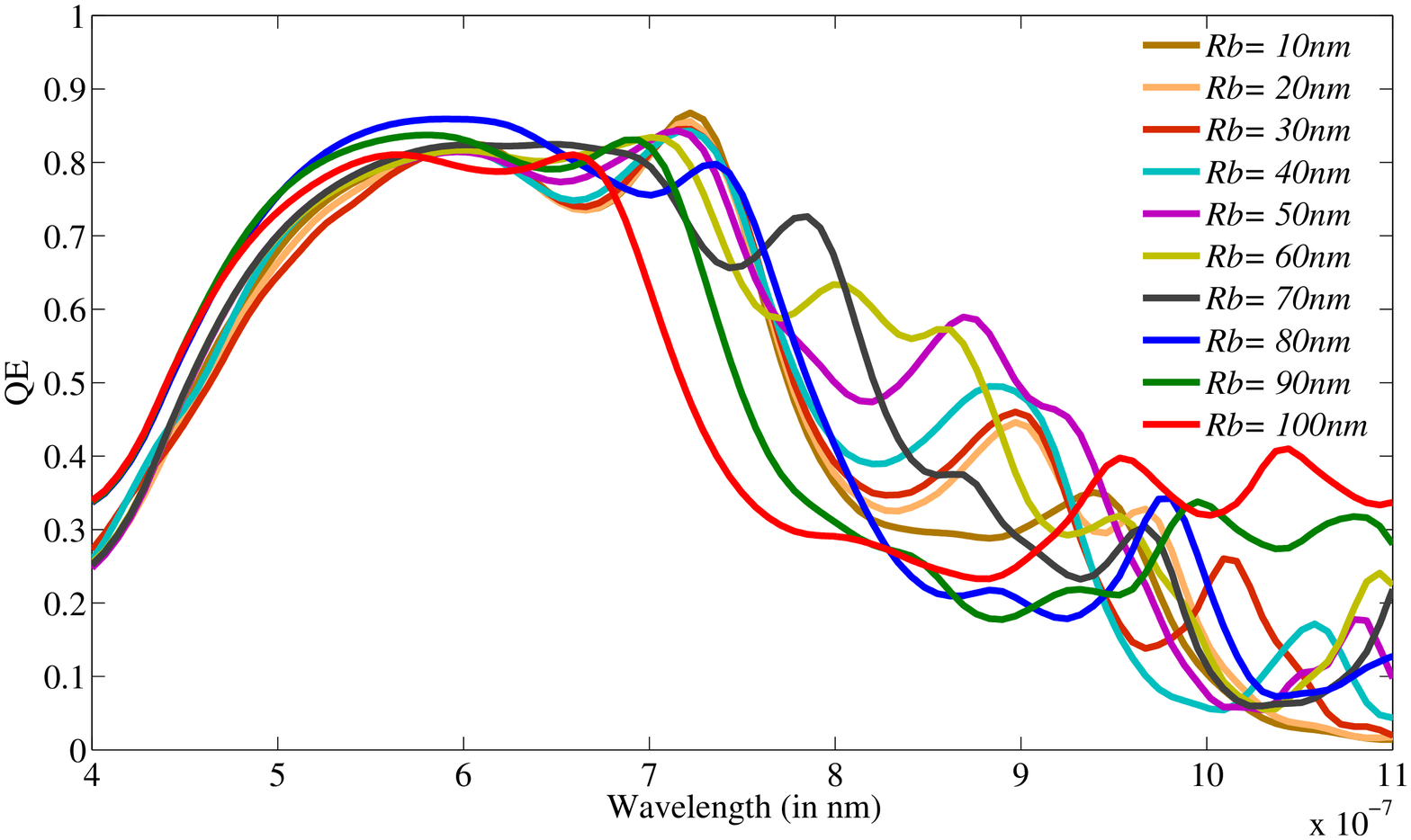}
  \caption{}
\end{subfigure}
\caption{Variation of $QE$ with wavelength for different (a)surface and (b)bulk layer particle radii.}
\end{figure}

Figure 5(a) shows that the absorption peak at wavelengths below $750\;nm$ shifts towards the right with increase in $R_s$. The magnitude of the absorption peak increases when the $R_s$ increases from $0$ to $60\;nm$. Further increase in $R_s$ results in the reduction of magnitude of absorption peak in this wavelength region. Another peak in $QE$ begins to appear at lower wavelengths around $400-550\;nm$ range for $R_s$ values higher than $70\;nm$. It is also observed that the $QE$ curve broadens as $R_s$ increases. 

The red-shift and broadening of $QE$ with increase in $R_s$ is because of dynamic depolarization and radiation damping~\cite{Meier,Kottman,Dahmen}. The field experienced by the nanoparticles can be considered as spatially constant but with time varying phase when the size of the homogeneous particle is much smaller than the wavelength of the incident light. This is known as the quasistatic limit\cite{Noguez}. For smaller values of $R_s$ the displacement of charges is homogeneous which result in a dipolar charge distribution on the surface. These charges give rise to only one proper resonance~\cite{Noguez}. As $R_s$ increases, the displacement of electronic cloud loses its homogeneity. As a result of this multipolar charge distributions are induced~\cite{Kreibig}. In addition to this, an additional polarization field is produced by the accelerated electrons~\cite{Meier83}. This field reacts against the quasistatic polarization field and shifts the position of the resonant modes to larger wavelengths. Also the electrons lose energy due to this secondary radiation and experience a damping effect which results in the widening of the range of frequencies over which surface plasmon resonance occurs~\cite{Noguez05}. Thus the increase in particle size  makes broader and asymmetric surface plasmon resonance peaks which are red-shifted and reduced in intensity.

Figure 5(b) shows that the particles in the bulk layer do not significantly affect the absorption at lower wavelengths but have a notable effect at higher wavelengths. The frequencies and intensities of localized surface plasmon resonances are known to be sensitive to the dielectric properties of the medium\cite{Haes,Hanarp,Jensen} and in particular, to the refractive index of matter close to the particle surface\cite{Haes,Jensen}. The larger particles show better absorption enhancement at higher wavelengths. The effect of varying nanoparticle radii of bulk and surface layer towards the absorption of photons over the entire spectrum is shown in Fig. 6. 

\begin{figure}[htb]
\centerline{\includegraphics[width=7cm]{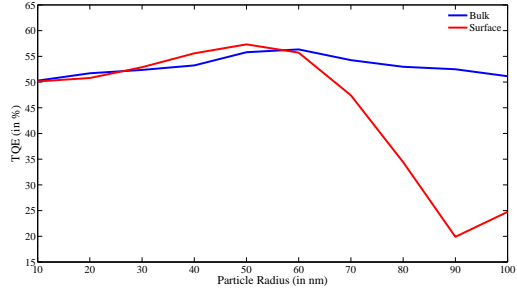}}
\caption{ Variation in $TQE$ with change in surface and bulk layer nanoparticle radii}
\end{figure}

It is evident from the graph that the surface layer radius variation affects the absorption to a higher extend. An initial increase in the radius of the surface layer nanoparticles causes an increment in $TQE$ but decreases when the particle radii exceeds $50\;nm$. Similarly, $TQE$ increases with increase in $R_b$ upto $60~nm$ and then decreases with further increment in $R_b$.  

\subsection{Influence of vertical alignment of particle layers}
The proposed design contains vertically aligned nanoparticle arrays in the surface as well as in the bulk of the a-Si substrate. But perfect vertical alignment is difficult to attain practically. In order to understand the extend to which the alignment of two nanoparticle layers affect the light absorption in the proposed design, the bulk array is moved in the horizontal direction while keeping the position of the surface array unaltered as shown in Fig. 7. The horizontal distance  $l$ between the centers of nanoparticles in both the layers is a measure of misalignment between the particle layers. 

\begin{figure}[h]
\centerline{\includegraphics[width=6.5cm]{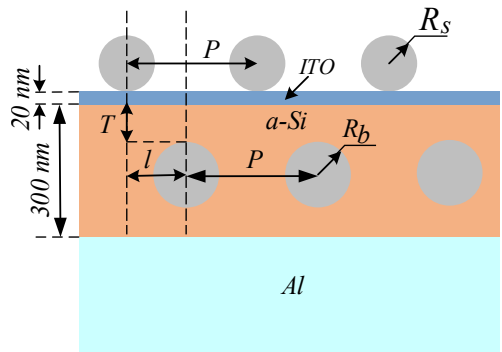}}
\caption{Alignment of particles in bulk and surface array.}
\end{figure}

The variation in absorption when the bulk array is moved away from the normal is shown in Fig. 8.

\begin{figure}[ht]
\centering
\begin{subfigure}{.48\textwidth}
  \centering
  \includegraphics[width=1\linewidth]{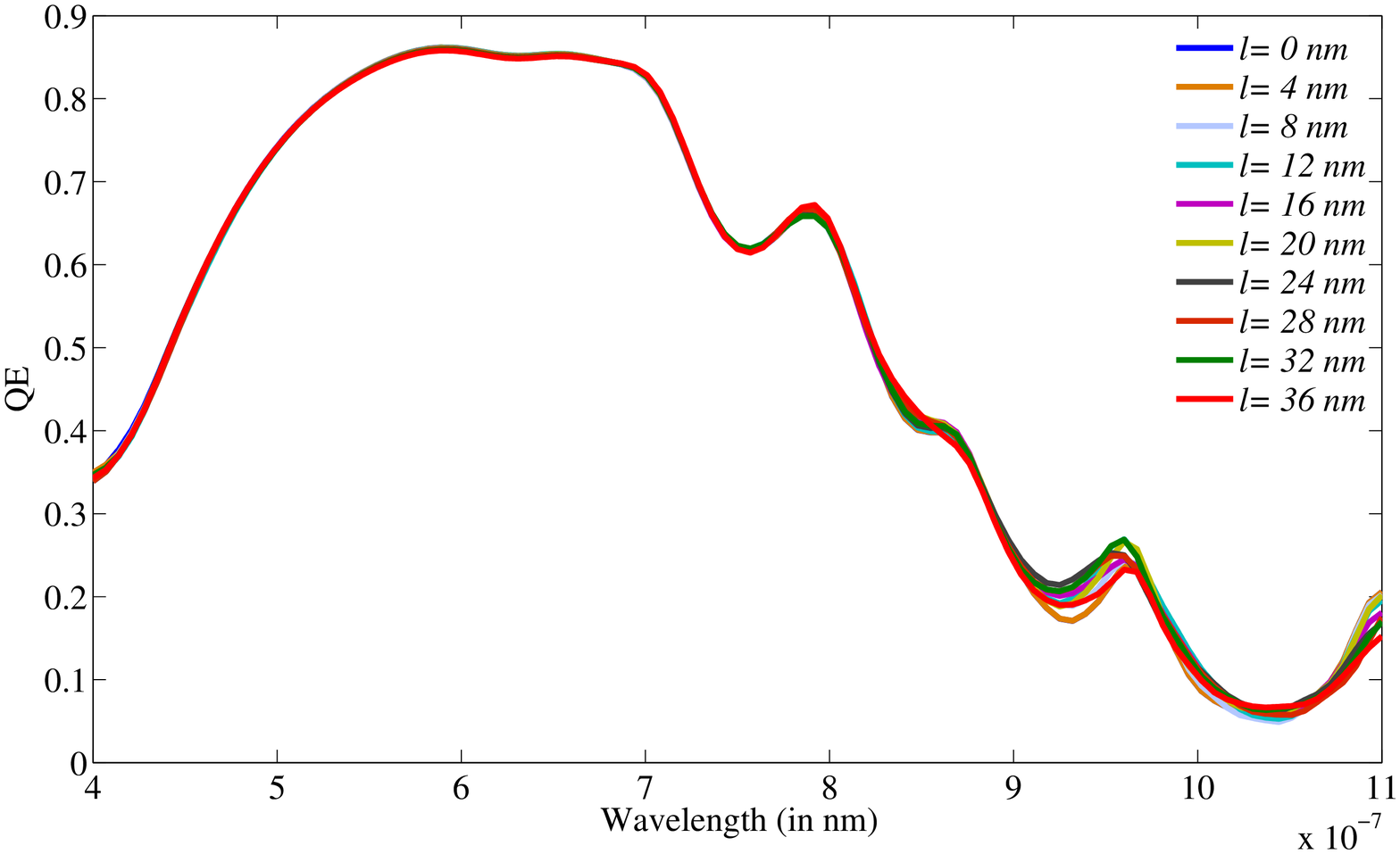}
  \caption{}
\end{subfigure}
\begin{subfigure}{.48\textwidth}
  \centering
  \includegraphics[width=1\linewidth]{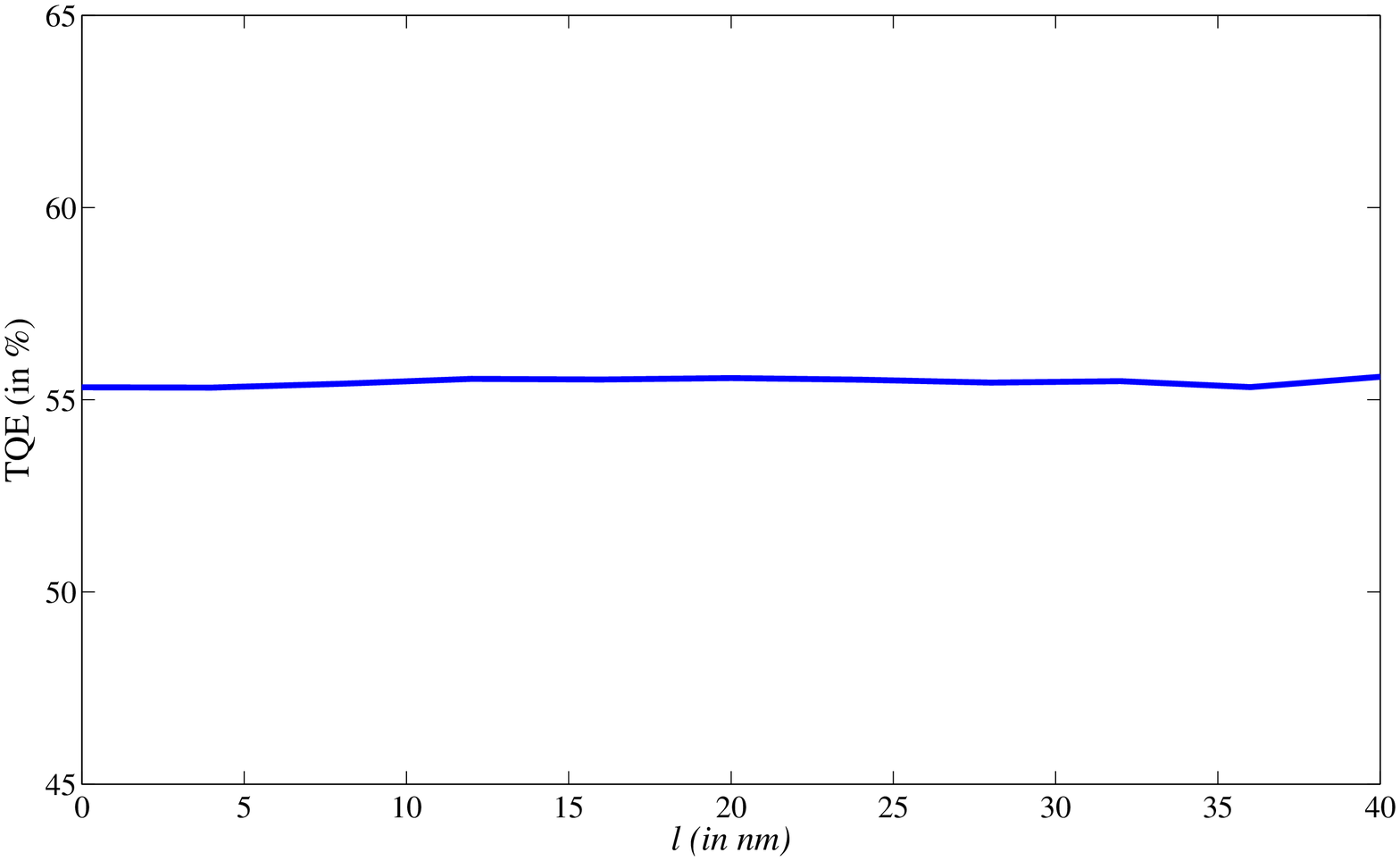}
  \caption{}
\end{subfigure}
\caption{Variation of $QE$ and $TQE$ with change in the alignment of particles.}
\end{figure}

The $QE$ and $TQE$ plot for various $l$ values shows that there is very little change in absorption efficiency when the nanoparticles in the bulk array is moved horizontally away from the normal passing through the center of the top layer nanoparticles. A slight mismatch in vertical alignment of nanoparticle arrays does not affect the overall absorption. Therefore the constraint of perfect vertical alignment can be relaxed while placing nanoparticle arrays in the solar cell.

\section{Optimization of design parameters}
The performance of the solar cell can be improved by tuning the parameters of the cell, namely the radii of the nanoparticles in both arrays($R_s,R_b$), the period of the rectangular array of particles ($P$) and the vertical distance between the two particle layers ($T$). The optimization is carried out using the Particle swarm optimization algorithm which converges to an optimum set of design parameters after a certain number of iterations~\cite{Robinson}. An a-Si active region of thickness $0.3~\mu m$ is considered for the solar cell. Values between $5\;nm$ and $100\;nm$ were considered for $R_s$ and $R_b$. The period, P, was considered from the minimum possible value, equal to the diameter of the larger particle among surface and bulk layer, to a maximum value of $600\;nm$. The value of $T$ is varied form zero to a maximum possible value equal to the difference between the thickness of the active region and diameter of bulk layer particle. 

\begin{figure}[h]
\centerline{\includegraphics[width=8.3cm]{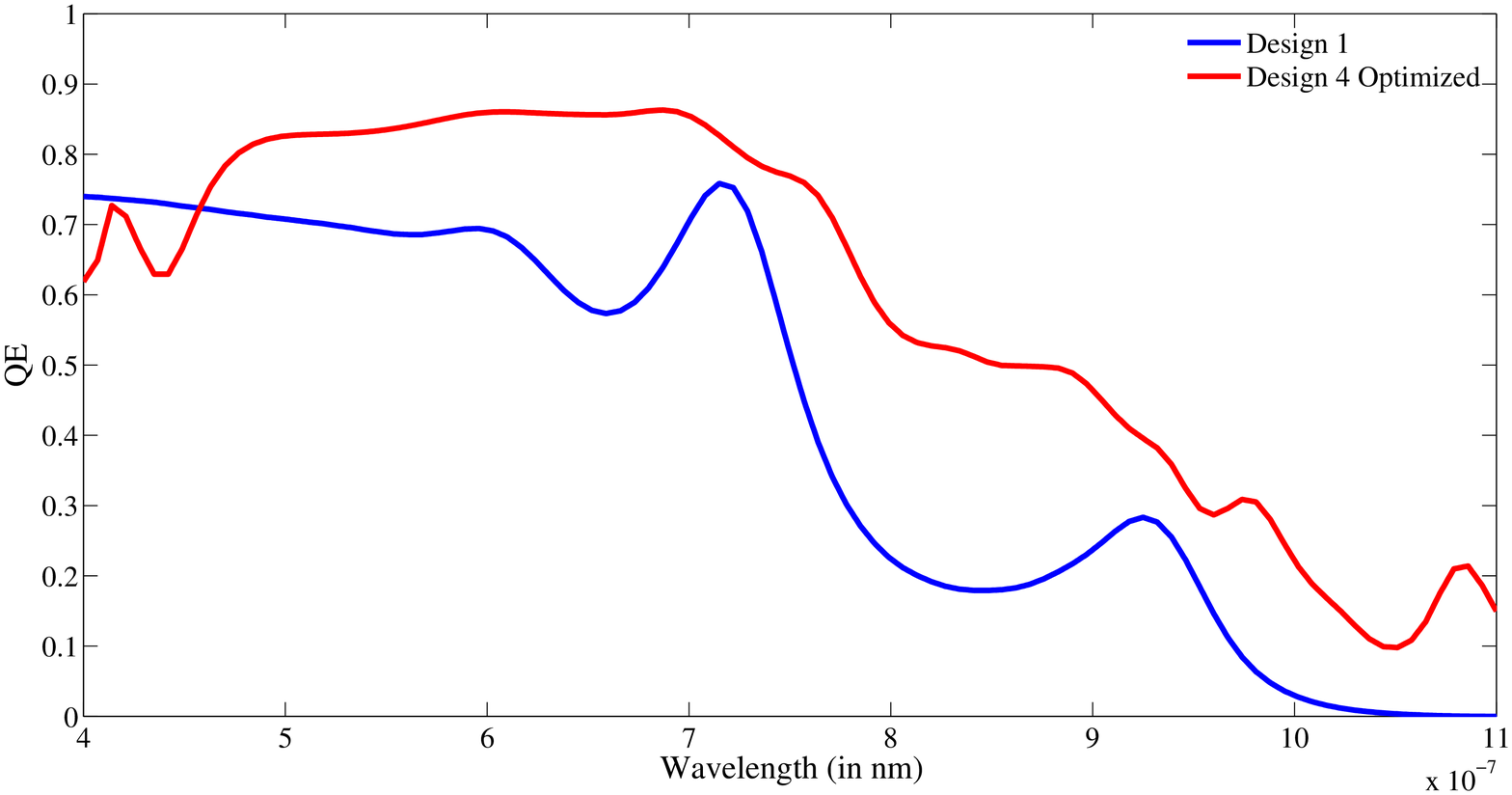}}
\caption{QE variation with wavelength for Design-1 and optimized Design 4}
\end{figure}

For the proposed solar cell design (Design 4), the optimal radii for the surface and bulk layer particles are found to be $87.30\;nm$ and $94.07\;nm$ respectively. A period of $413.22\;nm$ is obtained for the optimized design. A distance of $9.49\;nm$ from the surface of the a-Si substrate to the top of the bulk layer is found to be optimal for the solar cell. The optimized design resulted in a $TQE$ of $62.35\%$ which is $1.43$ times higher than the reference a-Si solar cell, Design 1. The quantum efficiency of the optimized design is compared with plain a-Si substrate in Fig. 9. The absorption efficiency of the substrate is improved for all wavelengths ranging from $450\;nm$ to $1100\;nm$ in the proposed design. 

\section{Conclusion}
The proposed solar cell design exploits the cumulative plasmonic enhancement from different nanoparticle layers at different wavelengths and improves the absorption efficiency over the entire range of wavelength. For calculating the short circuit current, recombination occurring in the device need to be considered. Recombination due to nanoparticles can be mitigated either by depositing nanoparticles at room temperature or by coating the nanoparticles with a thin dielectric layer. We studied the effect of dielectric coating thickness on $TQE$ and found that the variation is minimal if the thickness is below $3~ nm$. If the a-Si film thickness is thin, the recombination in the bulk of a-Si can be neglected~\cite{Pala}. Hence the enhancement in the short circuit current will be a replica of the enhancement in $TQE$. The photocurrent can be directly calculated from the absorption spectra of the solar cell~\cite{Pala, Deparis}.

In summary, the theoretical study shows that the thin film amorphous silicon solar cell with nanoparticle arrays placed on the surface and in the semiconductor layer showed significant improvement in absorption efficiency. The metallic nanoparticles at the surface contributes to the enhancement at  lower wavelengths and those embedded within the absorber layer contributes to the enhancement at higher wavelengths. The inter layer particle interaction further enhances the field intensity at intermediate wavelengths. Therefore, the design put forward in this paper exploits the surface plasmon resonance from metallic nanoparticles at all wavelengths and thus provides significant improvement in the overall light absorption efficiency.

\end{document}